\documentclass[prl,twocolumn,showpacs,superscriptaddress]{revtex4}
\usepackage{dcolumn}
\usepackage{graphicx}
\usepackage{amsmath}
\begin{document}
\title{Snell's law for surface electrons: Refraction of an electron gas imaged in real space}
\author{Jascha Repp}
\author{Gerhard Meyer}
\affiliation{
IBM Research, Zurich Research Laboratory, CH-8803 R\"uschlikon, Switzerland}
\affiliation{
Institut f\"ur Experimentalphysik, Freie Universit\"at Berlin, Arnimallee 14,
D-14195 Berlin, Germany }
\author{Karl-Heinz Rieder}
\affiliation{
Institut f\"ur Experimentalphysik, Freie Universit\"at Berlin, Arnimallee 14,
D-14195 Berlin, Germany }

\date{\today}

\begin{abstract}
On NaCl(100)/Cu(111) an interface state band is observed that
descends from the surface-state band of the clean copper surface.
This band exhibits a Moir\'e-pattern-induced one-dimensional band
gap, which is accompanied by strong standing-wave patterns, as
revealed in low-temperature scanning tunneling microscopy images.
At NaCl island step edges, one can directly see the refraction of
these standing waves, which obey Snell's refraction law.
\end{abstract}

\maketitle

Ever since the mapping of standing waves of a two-dimensional
electron gas (2DEG) on the close-packed Cu(111) surface by means
of low-temperature scanning tunneling microscopy (LT-STM)
\cite{Crommie93,Hasegawa93}, the 2DEG turned out to be an ideal
playground for a variety of LT-STM experiments
\cite{Buergi98,Ji03,Manoharan01,Kliewer01a,Suzuki01,Bendounan03,Diekhoener03,Kliewer01b,Park00,Repp00}.
In the STM images the surface electrons reveal their wave-like
behavior, often discussed in analogy to light. This analogy has
been addressed directly in experiments such as the confinement of
a 2DEG between two parallel step edges, considered as a
counterpart of the optical Fabry-Perot resonator~\cite{Buergi98},
the striking experiment of an electronic Mach-Zehnder
interferometer~\cite{Ji03}, and the spectacular quantum mirage
experiment~\cite{Manoharan01}. However, despite this strong
analogy, the counterpart of optical refraction has not yet been
observed for surface electrons.

Such an experiment has to comply with several requirements. First
of all, two different media are necessary. For surface electronic
states, two regions having different dispersions may act as the
media and be realized by a partial coverage of the surface with an
adlayer that modifies the dispersion, as was observed for various 
thin metal films on Cu(111)~\cite{Kliewer01a,Suzuki01,Bendounan03,Diekhoener03} 
as well as for Xe/Cu(111)~\cite{Park00}.
Second---in terms of optics---the interface between the two media
has to be transparent. It was reported that step edges of metals
exhibit a transmission close to zero for surface electrons, i.e.,
that the wave patterns on the two sides of a step edge are not
related to each other~\cite{Buergi98}. However, in the case of an
insulator adsorbed on a metal surface, this may be different: An
insulator does not contribute to the electronic states, therefore
the electrons are still confined to the surface of the metal
underneath, and thus may better overlap with the electrons of the
clean surface. Third, in addition to being inherently transparent,
the interface between the two media also has to be very smooth.
This means that a perfect step edge is needed to observe
refraction of surface electrons.

As NaCl/Cu(111) meets the above requirements, we are able to
report here the first observation of the refraction of surface
electronic waves at island step edges by means of LT-STM.
Moreover, we show that natural Moir\'e patterns, inherent in
incommensurate growth, generate a band gap within a
two-dimensional (2D) interface state band. Thereby the textbook
band-structure model of nearly free electrons (NFE) is exploited
and visualized in real space.


Our experiments were performed with a LT-STM~\cite{Meyer96}
operated at 9 K. The sample was cleaned by several sputtering and
annealing cycles. NaCl was evaporated thermally, while the sample
temperature was kept at 320 K. Bias voltages refer to the sample
voltage with respect to the tip, and we used electrochemically
etched tungsten wire as STM tips.

\begin{figure}
\centerline{\includegraphics[width=5cm]{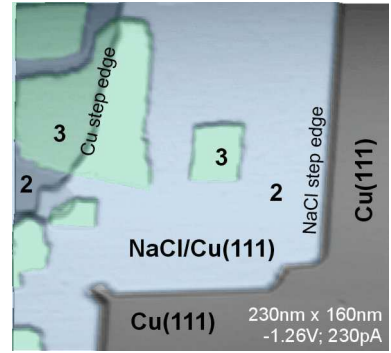}}
\caption{STM image of NaCl on Cu(111). The numbers indicate the
thickness of the NaCl islands in layers.}
\label{FigTopo}
\end{figure}

NaCl forms (100)-terminated islands that are up to several microns
in diameter~\cite{Bennewitz99} (see Fig.~\ref{FigTopo}). The NaCl
islands start with a double-layer thickness and perfect nonpolar
step edges, in which the anions and the cations alternate. On top
of the initial double layer, islands of additional layers are
formed. Substrate defect steps are smoothly overgrown in the
so-called carpet-like growth~\cite{Schwennicke93}. Because of the
different symmetries of NaCl(100) and Cu(111), the growth is
incommensurate. In the case of only a few layers of NaCl, the
nonvanishing tail of charge density from the metal that extends
through the insulator is sufficient to take STM images without
crashing the tip. In atomically resolved STM images, only the Cl
ions are imaged as protrusions~\cite{Hebenstreit99,Repp01,Olsson}.

\begin{figure}
\centerline{\includegraphics[width=7cm]{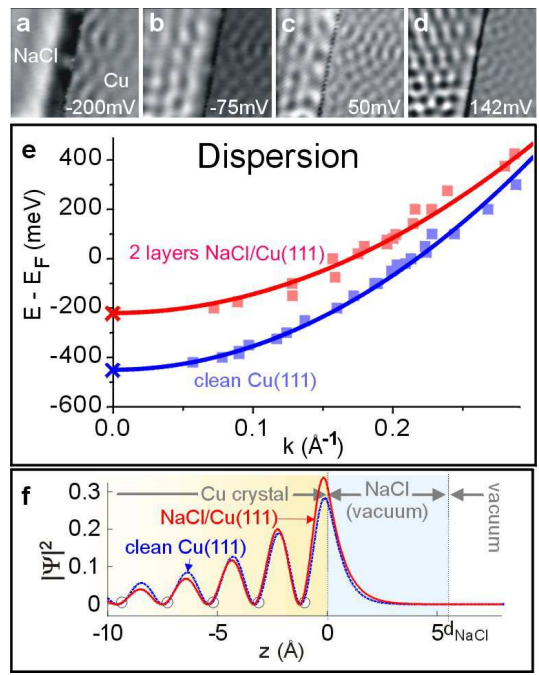}}
\caption{(a) to (d) d$I$/d$V$ images of the same region near a
NaCl step edge, taken at various voltages. In all cases the
wavelength of the NaCl/Cu(111) interface state is larger than that
of the Cu(111) surface state. (e) Dispersion of the 2DEG for
the clean Cu(111) surface and the NaCl(100)/Cu(111) interface
state band (square data points and fit curves). 
(f)
Using the phase-accumulation model, the probability distributions
$|\Psi|^2$ perpendicular to the surface are determined, yielding
only small changes upon the adsorption of NaCl. The origin ($z =
0$) was chosen to be the Cu(111) substrate surface. Circles denote
the position of Cu atoms. }
\label{FigDisp}
\end{figure}

Upon adsorption of NaCl, the Cu(111) surface state band survives
and forms an interface state band that is confined to the
insulator/metal interface and has a larger Fermi wavelength of
$\lambda_F = 3.8$ nm (clean Cu:\ $\lambda_F = 3.0$ nm). To measure
the dispersion of this modified 2D interface state band,
differential conductance (d$I$/d$V$) images were taken at various
bias voltages, as shown in Figs.~\ref{FigDisp}(a)-(d). By
extracting the wavelength of the wave patterns seen in these
images~\cite{constantcurrent}, the dispersion relation can be
established (see Fig.~\ref{FigDisp}(e)), as was done  for the
clean Cu surface~\cite{Crommie93,Hasegawa93,Horm94}. The band
minimum $E_0$ was obtained by recording the d$I$/d$V$ signal at a fixed
position as a function of energy~\cite{Horm94,Li98} (crosses in Fig.~\ref{FigDisp}(e)) and
is shifted upward in energy by
$(230\pm30)$ meV in the case of the adsorption of an insulating
NaCl overlayer.
The resulting
parabolic dispersion $E=E_0+(\hbar k)^2/2m^*$ is slightly wider,
i.e., the effective mass $m^*$ has increased  from $(0.40\pm0.02)m_e$ to
$(0.46\pm0.04)m_e$, where $m_e$ denotes the free electron mass.


The energy shift can be qualitatively
understood within the one-dimensional (1D) phase-accumulation
model~\cite{Smith85,Hotzel99}. In this analysis, the wave
function $\Psi$ of surface electrons is found by matching the
wave-function phase of the analytical solution inside the copper
crystal~\cite{Forstmann70} to the one of the numerical solution
for the outside region, obtained by integrating the Schr\"odinger
equation~\cite{Smith85}. For the latter an electrostatic potential
had to be assumed: For the clean Cu surface, this is given by the
image potential~\cite{potential}. For NaCl/Cu(111) we considered
the modified image potential~\cite{Hotzel99,potential} as well as the
lowering of the work function upon NaCl adsorption
\cite{Bennewitz99}.

This model yields an upward shift of the
dispersion of $\Delta E\simeq300$ meV, in qualitative agreement
with our experiment. Even more relevant, it reveals that the wave
functions of the surface electrons are barely modified upon
adsorption of NaCl (see Fig.~\ref{FigDisp}(f)) and thus render a
high transmission through NaCl step edges promising. Moreover, it
justifies an even simpler understanding of the upward shift in the
dispersion within first-order perturbation theory, $\Delta E =
\langle\Psi|\Delta V|\Psi\rangle$. Owing to the exponentially
decaying wave function $\Psi$ outside the Cu substrate, the
perturbation potential $\Delta V$, i.e., the potential difference
with and without NaCl, is only relevant for the energy shift
$\Delta E$ in immediate proximity of the copper surface. The upward shift
thus results from the positive perturbation potential at the
interface given by the weaker image potential \cite{potential} due to the
polarization of the adsorbed dielectric. The work-function change
\cite{Bennewitz99} as well as the image potential at the
insulator-to-vacuum interface are negligible for $\Delta E$.
Consequently, the dispersion for three layers of NaCl does not
differ from that for two layers, which was verified in the
experiment.

The above results show 
that for interface states of an insulator/metal system, as described here,
the main contribution of the electronic wavefunctions resides within 
the substrate (see Fig.~\ref{FigDisp}(f)).
Therefore, the effective mass of the shifted interface state
will mainly be determined by the metal.
For copper an upward shift of the surface state is expected 
to be correlated with a slight increase of the effective mass as discussed in Ref.~\cite{Smith85},
in good agreement with the present experiment.
The energetic shift of the band for other insulator/metal systems will depend critically on the 
structure of the interface.
However, the adsorption of an insulator that does not strongly affect 
the metal surface but weaken its
image potential will in general result in an upward shift of the 
surface state band.
These findings explain that in a previous experiment on Xe/Cu(111)
\cite{Park00} a similar upward shift and 
effective mass increase were observed.


\begin{figure}
\centerline{\includegraphics[width=6cm]{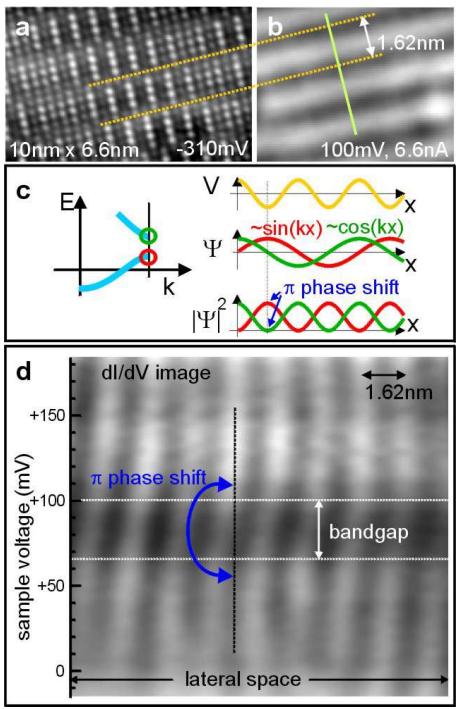}}
\caption{(a) An atomically resolved STM image shows one of the
various possible Moir\'e patterns that result from incommensurate
growth~\cite{pattern}. (b) This pattern gives rise to a strong
plane-wave pattern. (c) Illustration of a band gap in the NFE
model. Just below and just above the band gap (marked by circles),
the electronic wave functions are expected to be standing waves,
and the probability distributions $|\Psi|^2$ will have a phase
shift of $\pi$ with respect to each other. (d) This d$I$/d$V$
image (acquired in constant-height mode) maps the probability
distribution $|\Psi|^2$ in real space and energy. It shows the
behavior expected from the NFE model. }
\label{FigGap}
\end{figure}

Another condition for the observation of the electron refraction
at island edges is the presence of standing waves. Continuous
waves lead to a uniform probability distribution and thus prevent
the observation of refraction in the STM images, even if the
refraction took place. In the case of NaCl/Cu(111), strong
electronic plane waves are already inherent in the system: The
incommensurate growth gives rise to various Moir\'e patterns
\cite{pattern}, which can be seen in atomically resolved STM
images (Fig.~\ref{FigGap}(a)). The Moir\'e patterns will be
associated with displacements of the charged ions and are thus
accompanied by a modulation of the electrostatic potential seen by
the surface electrons. One general conclusion of the NFE model
\cite{Ashcroft} is that any periodic potential modulation will
give rise to a band gap. Just below and just above the band gap,
the electronic wave functions are expected to be standing waves,
and the probability distributions $|\Psi|^2$ will have a phase
shift of $\pi$ with respect to each other, as illustrated in Fig.~\ref{FigGap}(c). 
For the case discussed here, this physics
textbook statement can be imaged in real space. As a first
indication, STM images such as that in Fig.~\ref{FigGap}(b) show
strong standing plane waves that correspond to this particular
Moir\'e pattern in terms of wavelength and direction (dotted lines
in Fig.~\ref{FigGap}(a) and (b)).

\begin{figure}
\centerline{\includegraphics[width=5cm]{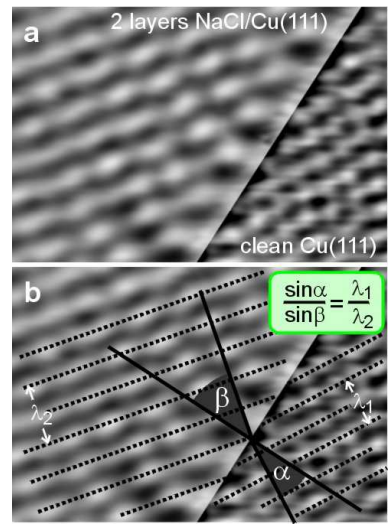}}
\caption{(a) STM d$I$/d$V$ image (constant-current mode) of a NaCl
island step edge reveals strong standing-wave patterns, not only
on the NaCl island but also on the clean Cu(111) surface ($V = 55$ mV). 
(b) Adding some lines as guide to the eye shows that the
standing-wave pattern obeys Snell's refraction law. Note that the
contrast for the area of clean Cu(111) is increased,
compared with the rest of the image.}
\label{FigRefr}
\end{figure}


For a more precise analysis, we recorded an d$I$/d$V$ image in
constant-height mode, but whereas we scanned within one scan line
perpendicular to the plane waves (as indicated by the solid line
in Fig.~\ref{FigGap}(b)), we did not move the tip in the other
direction. Instead, we changed the bias voltage after each scan
line. The resulting STM image in Fig.~\ref{FigGap}(d) thus
reflects a map of the probability distribution $|\Psi|^2$ in space
and energy. As expected, the image shows a region of reduced
intensity, which represents the band-gap region. Just above and
just below this region, there are strong standing-wave patterns
shifted by $\pi$ to each other.

The standing-wave pattern caused by a periodic scattering
potential contains only the wavelength components of this
potential modulation. Thus in Fig.~\ref{FigGap}(d), the
standing-wave pattern is given by the Moir\'e pattern's
periodicity and does not change with energy. 
The position in energy of the band gap can vary between 20 and 400 meV 
above the Fermi level from one rotational domain to the next.
In all cases, this energy position 
can be linked to the Moir\'e
patterns in such a way that the wave vector at the band gap 
corresponds to half of the wave vector of the potential
modulation, in agreement with the NFE model. 
In the particular case shown in Fig. \ref{FigGap}
the Moir\'e pattern has a periodicity of $\simeq 1.62$ nm corresponding to 
$k=2\pi/16.2$ \AA$^{-1}$, $k_{\rm gap}=0.194$ \AA$^{-1}$, and thus $E_{\rm gap}\simeq81$ meV.
As this condition is
met only in one direction, we have a 1D band gap in a 2D electron
gas, i.e., a pseudo gap. For some rotational domains, the condition was not met in the
experimentally accessible energy range, and no band gap was
observed, as in the case of the dispersion shown in Fig.~\ref{FigDisp}(e).

The standing-wave pattern persists far above
the band gap because of the one-dimensionality of the band gap. 
This is also one of the reasons for the remaining low intensity
within the band gap.
In addition, the finite lifetime of the interface states~\cite{Kliewer01b}
gives rise to an energetic broadening of the 
states, which further ``smears out'' the bandgap and may also
lead to interference effects.


To observe the refraction, we took advantage of the strong
standing plane waves just below the band gap by applying a
corresponding bias voltage of $V=55$ meV. Figure~\ref{FigRefr}
shows the refraction of the surface electrons in a differential
conductance (d$I$/d$V$) image of an island edge of NaCl/Cu(111).
The region on the NaCl island (to the left) shows a strong
standing plane-wave pattern that is due to the formation of a band
gap. Reflection induces an additional modulation of the pattern
parallel to the step edge. More importantly, the image clearly
reveals standing plane waves on the clean copper surface (to the
right). These patterns at the island step edge obey Snell's
refraction law, $\sin(\alpha)/\sin(\beta)=\lambda_1/\lambda_2$. 
Note that $\beta$ and $\lambda_2$ are given by the geometry of this
particular Moir\'e pattern, $\lambda_1$ is the electron wavelength
on the clean Cu surface corresponding to $E=E_F+55$ meV, and $\alpha$ 
is given by Snell's law.
This law follows directly from the conservation of
the component of the momentum (and wavelength) parallel to the
step. In other words, the patterns on both sides have a fixed
phase correlation at the step edge, which is directly evident in
the images and follows from the postulation of continuous wave
functions in quantum mechanics.

In everyday life, optical refraction manifests itself in the
bending of light rays, and we cannot observe the wave fronts
directly. In Fig.~\ref{FigRefr} we do not observe the bending of
rays, but directly image the wave fronts, visualizing the
refraction model. Another difference to optics is evident: In
optics the refraction is usually determined by one term only,
namely, the refractive index of the medium. In the case of
electron refraction, the band minimum as well as the effective
mass of the electrons may change and thus contribute to the
refraction in different ways.


The observed band-gap formation due to Moir\'e patterns is similar
to the superlattice concept~\cite{Esaki70,Dingle75} based on the
idea of creating artificial, tunable 1D band gaps by growing a
vertical lattice of alternating semiconductors. Whereas in
semiconductor superlattices the band gaps can be tuned by the
layer thicknesses during growth, Moir\'e-pattern-induced band gaps
can be tuned by the choice of lattice constant mismatch. This
phenomenon shows a new way to tailor the properties of a 2DEG for
future applications. These possibilities can be further extended
by controlled sequential growth of different dielectric materials.
In contrast to surface states, interface states
are inherently protected by the dielectric adlayer
and can even be studied under ambient conditions.


In summary, we observed the formation of a 1D band gap in the 2D
electron gas of the NaCl/Cu(111) interface state band. The band
gap is due to the Moir\'e patterns of the incommensurate growth
and was observed in an STM image displaying the
electron-probability distribution in space and energy. At NaCl
island edges, the refraction of standing waves could be observed,
obeying Snell's refraction law.

\begin{acknowledgements}
We thank Frank Forstmann and Rolf Allenspach for fruitful
discussions, and acknowledge partial funding by the EU-TMR projects
``AMMIST'' and ``NANOSPECTRA'' and the Deutsche Forschungsgemeinschaft Project No.\ RI
472/3-2.
\end{acknowledgements}


\begin{thebibliography}{99}
\bibitem{constantcurrent}
The d$I$/d$V$ images presented in Figs.~\ref{FigRefr} and
\ref{FigDisp} are recorded in constant-current mode to enable
simultaneous imaging on both sides of the step edge. To determine
the dispersion, constant-height images were used as well.
\bibitem{potential}
As in Ref. \cite{Hotzel99} we treated the image potential classically by
applying the static dielectric constant of NaCl.
The nonphysical divergence of the image potential was suppressed.
The average barrier potential within the NaCl layer is given by the upper
edge of the band gap.
\bibitem{pattern}
The pattern shown in Fig.~\ref{FigGap}(a) corresponds to a NaCl
lattice constant of $a = 5.52$ \AA\/ and a rotational domain in
which the polar NaCl direction is almost parallel
($\alpha=0.2^\circ$) to the close-packed direction of the copper
substrate. This alignment of the two directions seems
favorable, although also other rotational domains exist.
\end{thebibliography}
\end{document}